\documentclass[%
 reprint,
 amsmath,amssymb,
 aps,
]{revtex4-2}

\usepackage{graphicx}
\usepackage{dcolumn}
\usepackage{bm}

\begin{document}
\title{Electric Analog of Magnons in Order-Disorder Ferroelectrics}

\author{Ping Tang$^{1}$}
\author{Gerrit E. W. Bauer$^{1,2,3,4}$}
\affiliation{$^1$WPI-AIMR, Tohoku
University, 2-1-1 Katahira, Sendai 980-8577, Japan}
\affiliation{$^2$Institute for Materials Research, Tohoku University,
2-1-1 Katahira, Sendai 980-8577, Japan} 
\affiliation{$^3$Center for Science and Innovation in Spintronics (CSIS), Tohoku University, Sendai 980-8577, Japan}
\affiliation{$^4$Kavli Institute for Theoretical Sciences, University of the Chinese Academy of Sciences,
Beijing 10090, China}
\date{\today}

\begin{abstract}
We analyze the ``ferron" excitations in order-disorder ferroelectrics by a microscopic pseudo-spin model. We demonstrate that analogous to magnons, the quanta of spin waves in magnetic materials, ferrons carry both static and oscillating electric dipole moments, exhibit a Stark effect, and may be parametrically excited by THz radiation. The anti-crossing gap of the ferron-photon hybrid depends strongly on propagation direction and an applied static electric field. We predict ferron diffusion lengths that can reach centimeters, which implies efficient transport of electric polarization by temperature gradients. These properties suggest that ferroelectric materials may be useful for information technology beyond data storage applications.
\end{abstract}

\maketitle
\emph{Introduction.---}In ferroelectrics and magnets, electric and magnetic dipoles order below a Curie temperature ($T_{c}$), respectively. The collective excitations of ferroic orders are bosonic quasi-particles that carry energy, momentum, and magnetic or electric polarization. In insulating magnets, electrically or thermally excited spin waves or their quanta, magnons, transport magnetization over large distances without the Joule heating \cite{Kajiwara2010, Cornelissen2015, Shan2016, Ganzhorn2017, Lebrun2018}, making them attractive for low-power devices \cite{Chumak2015}. The magnon polariton, i.e. the hybridized state between magnons and photons, is a promising conduit for quantum information technologies \cite{Rameshti2022}. However, analogous elementary excitations in ferroelectrics have so far received little attention \cite{Bauer2022}.

Most ferroelectrics are of displacive or order-disorder type with phase transitions triggered by the condensation of a soft optical phonon or the ordering of electric charges \cite{Bruce1981}. We recently addressed the transport of heat and electric polarization in ferroelectrics in terms of quasi-particle excitations called \textquotedblleft ferrons\textquotedblright\ \cite{Bauer2021, Tang2022a} for displacive ferroelectrics \cite{Tang2022b} by the phenomenological Landau-Ginzburg-Devonshire theory \cite{Devonshire1949, Devonshire1951}. These ferrons emerge from anharmonic \emph{longitudinal} fluctuations of the ferroelectric order. On the other hand, the elementary excitations of order-disorder ferroelectrics can be modeled by the dynamics of pseudo-spins \cite{Gennes1963,Blinc1972,Blinc1974}. However, it remains an open question whether the associated pseudo-magnons (ferrons) can efficiently transport electric dipoles and interact with photons. 

Here we show that ferrons in order-disorder ferroelectrics indeed emerge from a microscopic pseudo-spin Hamiltonian. They carry both ac and dc electric dipole moments in the form of transverse fluctuations of the pseudo-spins. We predict the parametric excitation of ferrons by infrared radiation at a threshold power that depends on their dipole moment. An electric-field-tunable ferron-photon hybrid state or ferron polariton is characterized by an anti-crossing gap that depends strongly on the propagation direction. We also compute a ferron diffusion length in the range of tens of nanometers to centimeters. Our work opens up strategies to employ the untapped resources provided by ferroelectric materials in information processing.

\begin{figure}
\centering
\par
\includegraphics[width=7.6cm]{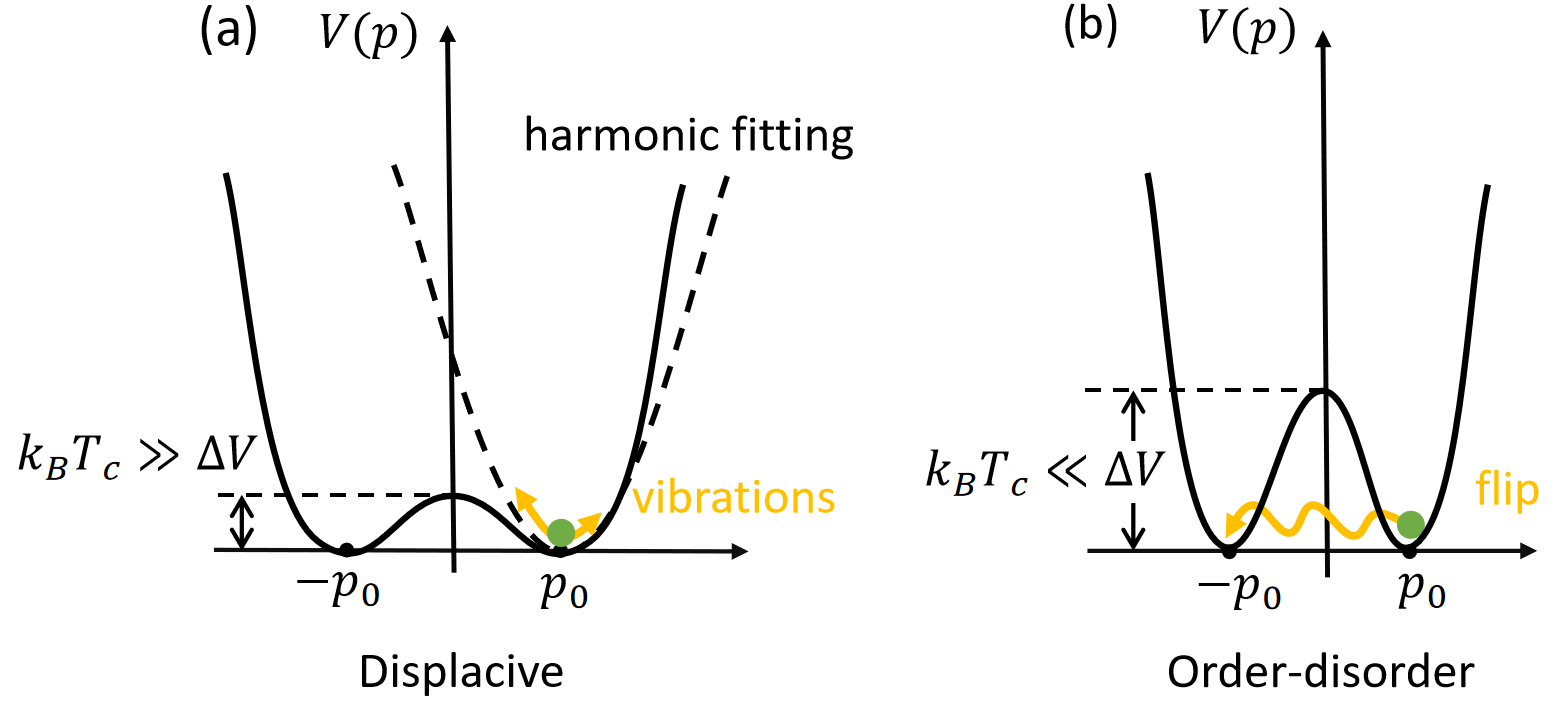}\newline \caption{The dynamics of the local (unit-cell) dipoles in a double-well potential for two types of ferroelectrics \cite{Bruce1981}.}%
\label{Fig-FE}%
\end{figure}

\emph{Pseudo-spin model and ferron excitations.---}The phase transition in most ferroelectrics can be described by the off-center dynamics of pertinent ions or charged molecular units in double-well single-cell potentials $V$ \cite{Stamenkovic1976, Stamenkovic1998}. Displacive and order-disorder transitions may be associated to the limiting forms of $V(p)$ as a function of the off-center displacement of the ions or the resultant unit-cell dipole moment \(p\) \cite{ Notedip}, as illustrated in Fig.~\ref{Fig-FE}. In displacive ferroelectrics, the ions and associated dipoles vibrate around their equilibrium positions with temperature-dependent amplitudes and the transition occurs at $k_{B}T_{c}\gg\Delta V$, where $\Delta V$ is the barrier height. In contrast, the potential barrier in order-disorder ferroelectrics remains $\Delta V\gg k_{B}T_{c}$ and the ferroelectricity weakens by a flip of the dipoles between the two polar states $\pm p_{0}$. The latter can be efficiently captured by the transverse Ising pseudo-spin Hamiltonian \cite{Gennes1963, Blinc1972, Blinc1974},
\begin{equation}
\hat{\mathcal{H}}=-\Omega \sum_{i}\hat{S}_{i}^{(x)}-\frac{1}{2}\sum_{i\neq
j}J_{ij}\hat{S}_{i}^{(z)}\hat{S}_{j}^{(z)}-2p_{0} E\sum_{i}\hat{S}_{i}^{(z)},  \label{HS}
\end{equation}
where $\hat{S}_{i}^{(\alpha)}$ is the $\alpha$-th Cartesian component of the pseudo-spin $S=1/2$ operator $\hat{\mathbf{S}}_{i}$ at site $i$ with up- and down-spin eigenstates of $\hat{S}_{i}^{z}$, $J_{ij}$ the coupling constant between the dipoles at different sites, $\Omega$ a transverse field, and $E$ an external uniform, static electric field along the polar ($z$) axis. In hydrogen-bonded ferroelectrics such as KH\(_2\)PO\(_4\) \cite{Blinc1972}, $\Omega$ parameterizes the proton tunneling or hopping between two energy minima of the O-H-O bond. 

We seek the state that minimizes Eq.~(\ref{HS}) in a mean-field approximation in a translationally periodic lattice in reciprocal space. To this end, the pseudo-spin order should maximize $J_{\mathbf{q}}=\sum_{\mathbf{r}_{ij}}J_{ij}\exp(-i\mathbf{q}\cdot\mathbf{r}_{ij})$, where $\mathbf{r}_{ij}$ is the displacement connecting $i$th and $j$th lattice sites. Here we adopt a model with both nearest-neighbor exchange-like and long-range dipolar interactions \cite{Yamada1966,Dumont1975}
\begin{equation}
J_{ij}=\frac{J_{0}}{z_{0}} F(r_{ij}) -\frac{(2p_{0})^2}{4\pi\varepsilon_{b}}\left(\frac{1}{r_{ij}^{3}}-\frac{3z_{ij}^3}{r_{ij}^{5}} \right) \label{Jij}
\end{equation}
where $F(r_{ij})=1$ for the nearest neighbor sites and otherwise $F(r_{ij})=0$, $z_{0}$ is the coordination number, and $\varepsilon_{b}$ the background dielectric constant. The exchange-like term may arise from the lattice deformations caused by the motion of active ions in neighboring cells \cite{Stamenkovic1976}. Assuming a simple cubic lattice and replacing the dipolar sum by an integral \cite{Note1}, we have $J_{\mathbf{q}}=(J_{0}/3)\left( \cos q_{x} a+\cos q_{y}a+\cos q_{z}a\right)-\hbar\omega_{D}\cos^2\varphi_{\mathbf{q}}$, where $a$ is the lattice constant, $\hbar\omega_{D}=4p_{0}^2/(\varepsilon_{b}a^{3})$ the energy scale of the dipolar interaction, and $\varphi_{\mathbf{q}}$ the angle of $\mathbf{q}$ relative to the polar axis. When $J_{0}>0$ the ground state is a uniform ferroelectric in which the pseudo-spins align along the direction of the molecular field $\mathbf{h}=( \Omega
,0,2p_{0} E+J_{0}\langle \hat{S}_{i}^{z}\rangle )$, where $\langle\cdots\rangle$ denotes an ensemble average. 

The quasi-particle excitations of Eq.~(\ref{HS}) are small-amplitude fluctuations around the equilibrium pseudo-spin order that is canted by an angle $\theta=\arcsin (\Omega/\vert\mathbf{h}\vert)$ from the $z$-axis. In terms of the $\hat{S}_{i}^{(\alpha\prime)}$ in a rotated coordinate frame with $z^{\prime}$ axis along the molecular field  
\begin{equation}
\left(\begin{matrix}
\hat{S}_{i}^{(x)}\\
\hat{S}_{i}^{(y)}\\
\hat{S}_{i}^{(z)}
\end{matrix}\right)=\left(\begin{matrix}
\cos\theta &0&\sin\theta\\
0& 1& 0\\
-\sin\theta & 0 &\cos\theta
\end{matrix}\right) \left(\begin{matrix}
\hat{S}_{i}^{(x\prime)}\\
\hat{S}_{i}^{(y\prime)}\\
\hat{S}_{i}^{(z\prime)}
\end{matrix}\right),
\end{equation}
in which $\langle \hat{S}
_{i}^{(x\prime)}\rangle =\langle \hat{S}_{i}^{(y\prime)}\rangle =0$. At sufficiently low temperatures, the fluctuations around the equilibrium are small and well described by the lowest-order expansion of the Holstein-Primakoff transformation of the rotated pseudo-spin operators: $\hat{S}_{i}^{(x\prime)}+i\hat{S}_{i}^{(y\prime)}=\sqrt{(1-\hat{a}_{i}^{\dagger}\hat{a}_{i})}\hat{a}_{i}\approx\hat{a}_{i}$, $\hat{S}_{i}^{(x\prime)}-i\hat{S}_{i}^{(y\prime)}=\hat{a}_{i}^{\dagger}\sqrt{(1-\hat{a}_{i}^{\dagger}\hat{a}_{i})}\approx\hat{a}_{i}^{\dagger}$ and $\hat{S}_{i}^{(z\prime)}=1/2-\hat{a}_{i}^{\dagger }\hat{a}_{i}$, where $\hat{a}_{i}$ and $\hat{a}%
_{i}^{\dagger }$ are bosonic annihilation and creation operators, respectively. With $\hat{a}_{\mathbf{q}}=(1/\sqrt{N})\sum_{i}\hat{a}_{i}e^{-i\mathbf{q}%
\cdot \mathbf{r}_{i}}$ and its Hermitian conjugate, where $N$ is the number of unit cells, we then arrive at an ensemble of harmonic oscillators:
\begin{align}
\hat{\mathcal{H}}^{(2)}= & \vert\mathbf{h}\vert\sum_{\mathbf{q}}
\hat{a}_{\mathbf{q}}^{\dagger }\hat{a}_{\mathbf{q}}-\frac{\sin^2\theta}{8}\sum_{\mathbf{q}}J_{\mathbf{q%
}}( \hat{a}_{\mathbf{q}}+\hat{a}_{-\mathbf{q}}^{\dagger }) ( 
\hat{a}_{-\mathbf{q}}+\hat{a}_{\mathbf{q}}^{\dagger })\nonumber\\
=&\sum_{\mathbf{q}}\hbar \omega _{\mathbf{q}}\hat{b}_{%
\mathbf{q}}^{\dagger }\hat{b}_{\mathbf{q}}+\text{const.}, \label{FHM}
\end{align}
where $\hbar\omega_{\mathbf{q}}=\vert\mathbf{h}\vert\sqrt{1-\gamma_{\mathbf{q}}} $ is the ferron energy dispersion with $\gamma_{\mathbf{q}}=\sin^2\theta J_{\mathbf{q}}/(2\vert \mathbf{h}\vert)$. The second equality follows from the Bogoliubov transformation $\hat{a}_{\mathbf{q}}=u_{\mathbf{q}}\hat{b}_{\mathbf{q}} +v_{\mathbf{q}}\hat{b}_{-\mathbf{q}}^{\dagger}$, with $u_{\mathbf{q}}=[(1-\gamma_{\mathbf{q}}/2)(1-\gamma_{\mathbf{q}})^{-1/2}+1]^{1/2}/\sqrt{2}$ and $v_{\mathbf{q}}=[(1-\gamma_{\mathbf{q}}/2)(1-\gamma_{\mathbf{q}})^{-1/2}-1]^{1/2}/\sqrt{2}$. The higher-order terms in the Holstein-Primakoff expansion correspond to ferron interactions that, among other effects, cause finite lifetimes as discussed below. 

The electric polarization in terms of the field operators \(\hat{a}\) and \(\hat{a}^{\dagger}\) reads
\begin{align}
  \hat{P}(\mathbf{r}, t)=&\frac{2p_{0}N}{V}\hat{S}_{i}^{(z)}\approx \frac{p_{0}N}{V}\left[1-\frac{\sin\theta}{\sqrt{N}}\sum_{\mathbf{q}}\left(\hat{a}_{\mathbf{q}}(t)e^{i\mathbf{q}\cdot\mathbf{r}}\right.\right.\nonumber\\
  &\left.\left.+\text{H.c.}\right)-\frac{2\cos\theta}{N}\sum_{\mathbf{k}\mathbf{q}}\hat{a}_{\mathbf{k}}^{\dagger}\hat{a}_{\mathbf{k}+\mathbf{q}}(t)e^{i\mathbf{q}\cdot\mathbf{r}}\right]  \label{Pol}
\end{align}
where $V$ is the system volume. In the presence of coherent ferron excitations with  $\langle\hat{b}_{\mathbf{q}}\rangle=\vert b_{\mathbf{q}}\vert\exp(-i\omega_{\mathbf{q}}t+i\phi_{\mathbf{q}})$, the polarization deviates from the ground state by
\begin{align}
\Delta &P=\frac{2p_{0}\sqrt{N}}{V}\sin\theta(u_{\mathbf{q}}+v_{\mathbf{q}}) \vert b_{\mathbf{q}}\vert \cos(\mathbf{q}\cdot\mathbf{r}-\omega_{\mathbf{q}}t+\phi_{\mathbf{q}})\nonumber\\
+& \frac{\delta p_{\text{ac}}}{V} \vert b_{\mathbf{q}}\vert^2\cos\left[2\left(\mathbf{q}\cdot\mathbf{r}-\omega_{\mathbf{q}}t+\phi_{\mathbf{q}}\right)\right]+\frac{\delta p_{\text{dc}}}{V}\vert b_{\mathbf{q}}\vert^2 \label{Pol2}
\end{align}
where $\vert b_{\mathbf{q}}\vert $, $\phi_{\mathbf{q}}$, and $\vert b_{\mathbf{q}}\vert^2$ are the amplitude, phase, and ferron number, respectively. The first two terms on the right-hand side oscillate in time with angular frequencies $\omega_{\mathbf{q}}$ and $2\omega_{\mathbf{q}}$, respectively. Only the last (dc) component survives a time average and remains finite for thermal (incoherent) ferrons with random phases. $\delta p_{\text{ac}}=-4p_{0} u_{\mathbf{q}}v_{\mathbf{q}}\cos\theta$ and $\delta p_{\text{dc}}=-2p_{0}(u_{\mathbf{q}}^{2}+v_{\mathbf{q}}^{2})\cos\theta$ are defined as the ac and dc electric dipole moments of a single ferron with wave vector \textbf{q}, respectively. Since $u_{\mathbf{q}}^{2}-v_{\mathbf{q}}^{2}\equiv 1$, $
\delta p_{\text{ac}}=-\sqrt{\delta p_{\text{dc}}^2-(2p_{0}\cos\theta)^2}$, i.e., the ac dipole moment is necessarily smaller than the dc one. 

\begin{figure}
\centering
\par
\includegraphics[width=7.6 cm]{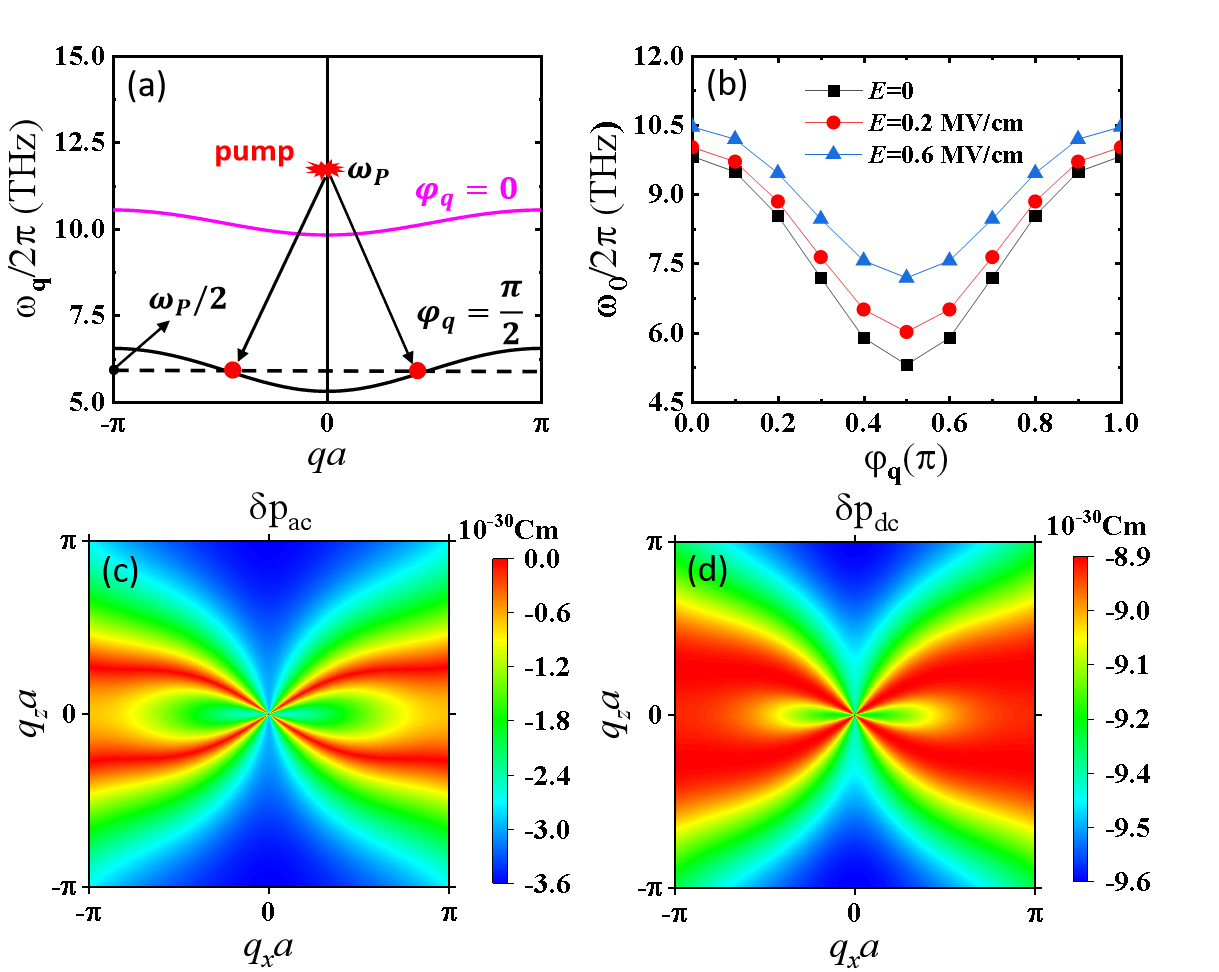}\newline\caption{(a) The top (bottom) ferron band for wave vectors parallel (perpendicular) to the polar ($z$) axis using the parameters in the text. The arrows illustrate the parametric pumping of ferron pairs by THz radiation with frequency $\omega_{P}$ with opposite wave vectors at half the frequency $\omega_{P}/2$. (b) The static electric field control of the propagation-direction dependent ferron gap or ferron Stark effect. (c) [(d)] The ac [dc] dipole moments of ferrons in reciprocal space. The wave vector is set in the $xz$ plane $\mathbf{q}=q(\sin\varphi_{\mathbf{q}},0, \cos\varphi_{\mathbf{q}})$ by symmetry, with \(\varphi_{\mathbf{q}}\) the angle with respect to the $z$ axis.}
\label{Fig-dispersion}%
\end{figure}

Fig.~\ref{Fig-dispersion} plots the ferron dispersion as well as wave vector dependent ac and dc dipole moments for the model parameters $\Omega=19.4\,$meV, $J_{0}=58.6\,$meV, $p_{0}=5.94\times10^{-30}\,$Cm, and $\omega_{D}/2\pi=43.9\,$THz of the order-disorder ferroelectric RbH$_2$PO$_4$ \cite{Peercy1974}. The dipolar interaction leads to an anisotropic dispersion that depends strongly on the wave vector direction relative to the polar axis. The dc dipoles emerge by the reduced average projection of the pseudo spin on the polar axis. Accordingly, an external static electric field tunes the ferron gap and enables the steering of excited states [Fig.~\ref{Fig-dispersion}(b)]. The ac dipoles proportional to the coherence factors $ u_{\mathbf{q}}v_{\mathbf{q}}$ reflect an oscillating projection by a canting of the pseudo-spin order relative to the polar axis. The ferron dispersion and dipoles are a necessary input for many observables of an excited ferroelectric. In the following, we discuss two examples, \textit{viz}. the interaction of coherent ferrons with ac electromagnetic fields in the THz regime and the non-equilibrium polarization and heat transport under temperature gradients.

\emph{Parametric excitation of coherent ferrons.---}An ac electric field can resonantly excite and be absorbed by optical phonons, but only ferrons can be excited parametrically at twice the resonance frequency. We shall show that the threshold excitation field provides direct information about the oscillating dipole moments \(\delta p_{\text{ac}}\). Consider electromagnetic THz radiation with a uniform ac electric field of frequency $\omega_{P}$ polarized along the polar axis $\mathcal{E}(t)=[\mathcal{E}_{0}\exp(-i\omega_{P}t)+\mathrm{H.c.}]/2$. The ferrons are ``pumped" by the dynamic Stark interaction 
\begin{align}
\hat{\mathcal{H}}&_{\text{pump}}=\frac{\sqrt{N}p_{0}}{2}\sin\theta\sum_{\mathbf{q}}(u_{\mathbf{q}}+v_{\mathbf{q}})\hat{b}_{\mathbf{q}}^{\dagger} \delta_{\mathbf{q},\mathbf{0}} \mathcal{E}_{0} e^{-i\omega_{P}t}\nonumber\\
&-\frac{1}{4}\sum_{\mathbf{q}}\delta p_{\text{ac}}(\mathbf{q})\hat{b}_{\mathbf{q}}^{\dagger}\hat{b}_{-\mathbf{q}}^{\dagger}\mathcal{E}_{0}e^{-i\omega_{P}t}+ \text{H.c.} \label{Pump}
\end{align}
The first term and its Hermitian conjugate represent the resonant light absorption, while the second describes the \textit{non-linear} event of creating two ferrons with opposite wave vectors and, by energy conservation, a frequency $\omega_{P}/2$. 

The amplitude of the ferron pair is governed by the Heisenberg equation of motion
\begin{equation}
i\hbar\frac{\partial\tilde{b}_{\mathbf{q}}}{\partial t}=\hbar(\delta\omega_{\mathbf{q}}-i\Gamma_{\mathbf{q}})\tilde{b}_{\mathbf{q}}-\frac{1}{2} \mathcal{E}_{0}\delta p_{\text{ac}} \tilde{b}_{-\mathbf{q}}^{\ast}, \label{feronpum1}
\end{equation}
 where $\delta\omega_{\mathbf{q}}=\omega_{\mathbf{q}}-\omega_{P}/2$ is the frequency detuning from the 2-ferron resonance and $\Gamma_{\mathbf{q}}$ is a phenomenological relaxation rate. Here we switched to reduced amplitudes denoted by the tilde, \textit{i.e.},  $\tilde{b}_{\mathbf{q}}=\langle \hat{b}_{\mathbf{q}} \rangle\exp(i\omega_{P}t/2) $ and $b_{-\mathbf{q}}^{\ast}=\langle \hat{b}_{-\mathbf{q}}^{\dagger}\rangle\exp(-i\omega_{P}t/2)$. Eq.~(\ref{feronpum1}) and its complex conjugate determine the amplitude of the ferron pair.
 
The ansatz solution $\tilde{b}_{\mathbf{q}}$, $\tilde{b}_{-\mathbf{q}}^{\ast}\propto e^{\alpha t}$ leads to $\alpha=-\Gamma_{\mathbf{q}}+[(\mathcal{E}_{0}\delta p_{\text{ac}})^2/(2\hbar)^2-(\delta\omega_{\mathbf{q}})^{2}]^{1/2}$. The ferron amplitude therefore grows exponentially in time ($\alpha>0$) when the pump field exceeds $\mathcal{E}_{\text{cr}}=2\hbar/\vert\delta p_{\text{ac}}\vert(\Gamma_{\mathbf{q}}^2+\delta\omega_{\mathbf{q}}^2)^{1/2}$, which is smallest for $\omega_{\mathbf{q}}=\omega_{P}/2$ with
 \begin{equation}
    \mathcal{E}_{\text{cr}}=\frac{2\hbar}{\vert\delta p_{\text{ac}}\vert}\text{min}\left.\left\{\Gamma_{\mathbf{q}}\right\}\right\vert_{\omega_{\mathbf{q}}=\omega_{P}/2} .\label{Cri}
 \end{equation}
Eq.~(\ref{Cri}) implies that the ferron pair with wave vector that minimizes $\Gamma_{\mathbf{q}}$ on the frequency shell $\omega_{\mathbf{q}}=\omega_{P}/2$ is the first one to become unstable (i.e., parametrically excited), with integrated absorption power $\sim \Gamma_{\mathbf{q}}\vert b_{\mathbf{q}}\vert^2$. This mode dominates the subsequent dynamics in which higher-order non-linearities regulate the divergence \cite{Rezendebook}. 

Propagating ferrons emit ac electric stray fields that, analogous to magnons \cite{Bracher2017,Chen2019}, can be detected non-locally by two inductive antennas on a thin ferroelectric film. Here we suggest a three-terminal device with a THz line source at the center and two detector strips on opposite sides that can identify the pumped ferrons, e.g., by time-correlation spectroscopy \cite{Sheng2023}. For frequency-dependent $\Gamma_{\mathbf{q}}=10^{-3}\omega_{\mathbf{q}}$, for example, $\mathcal{E}_{\text{cr}}\sim 100\,$kV/cm at $\omega_{P}/2\pi=10\,$THz. On the other hand, when $\Gamma_{\mathbf{q}}$ is known, \textit{e.g.} from Raman scattering experiments, the ac ferron dipole moment \(\vert\delta p_{\text{ac}}\vert\) at half-frequency of the pump field can be inferred from Eq.~(\ref{Cri}) via the threshold field.

\emph{Hybridization of ferrons and photons.---}---Next, we address the coupling of a bulk ferron excitation with an ac electric field, in which a new hybridized state--the ferron polariton--emerges. The dynamics of electric polarization $\partial_{t}\mathbf{P}(\mathbf{r},t)$ emits electromagnetic fields that in the strong coupling regime must be treated self-consistently. In linear response, the amplitude of the excited ferrons reads 
\begin{equation}
b_{\mathbf{q}}(t)=\frac{\sqrt{N} p_{0}\sin\theta}{V}\int \frac{d\omega}{2\pi}\frac{(u_{\mathbf{q}}+v_{\mathbf{q}})e^{-i\omega t}}{\hbar(\omega-\omega_{\mathbf{q}}+i\Gamma_{\mathbf{q}})}\mathcal{E}_{z}^{(p)}(\mathbf{q},\omega) \label{amplitude}
\end{equation}
where $\mathcal{E}_{z}^{(p)}(\mathbf{q},\omega)$ is the Fourier component of the radiated electric field $\boldsymbol{\mathcal{E}}^{(p)}(\mathbf{r},t)$ along the polar axis. From the Maxwell equation $\boldsymbol{\nabla}\times(\boldsymbol{\nabla}\times \boldsymbol{\mathcal{E}}^{(p)})+(1/c^2)\partial_{t}^2\boldsymbol{\mathcal{E}}^{(p)}=-\mu_{0}\partial_{t}^2\mathbf{P}$ and $\boldsymbol{\nabla}\cdot(\boldsymbol{\mathcal{E}}^{(p)}+\varepsilon_{b}\mathbf{P})=0$, where $c=1/\sqrt{\varepsilon_{b}\mu_{0}}$ is the light velocity and $\mu_{0}$ the magnetic permeability, it follows that
\begin{align}
\boldsymbol{\mathcal{E}}^{(p)}(\mathbf{r},t)=&\frac{\sqrt{N} p_{0}\sin\theta}{\varepsilon_{b} V}\sum_{\mathbf{q}} \frac{(u_{\mathbf{q}}+v_{\mathbf{q}}) e^{i\mathbf{q}\cdot\mathbf{r}}}{\omega^2-c^2\mathbf{q}^2}b_{\mathbf{q}}(t) \nonumber\\
&\times \left(\omega^2\hat{\mathbf{z}}-c^2q_{z}\mathbf{q}\right)+\text{H.c.}. \label{Eletro}
\end{align}
Combining Eq.~(\ref{amplitude}) and Eq.~(\ref{Eletro}) leads to the ferron polariton dispersion relation $\omega_{\mathbf{q}\pm}=\sqrt{\mathcal{B}_{\mathbf{q}}\pm\sqrt{\mathcal{B}_{\mathbf{q}}^2-4\mathcal{C}_{\mathbf{q}}}}/\sqrt{2}
$, where $\mathcal{B}_{\mathbf{q}}=\omega_{\mathbf{q}}^2+c^2\mathbf{q}^2+(1/2) (u_{\mathbf{q}}+v_{\mathbf{q}})^2\omega_{D}\omega_{\mathbf{q}}\sin^2\theta$ and $\mathcal{C}_{\mathbf{q}}=c^2\mathbf{q}^2[\omega_{\mathbf{q}}^2+(1/2) (u_{\mathbf{q}}+v_{\mathbf{q}})^2\omega_{D}\omega_{\mathbf{q}}\sin^2\theta]$. 
\begin{figure}
\centering
\par
\includegraphics[width=7.6cm]{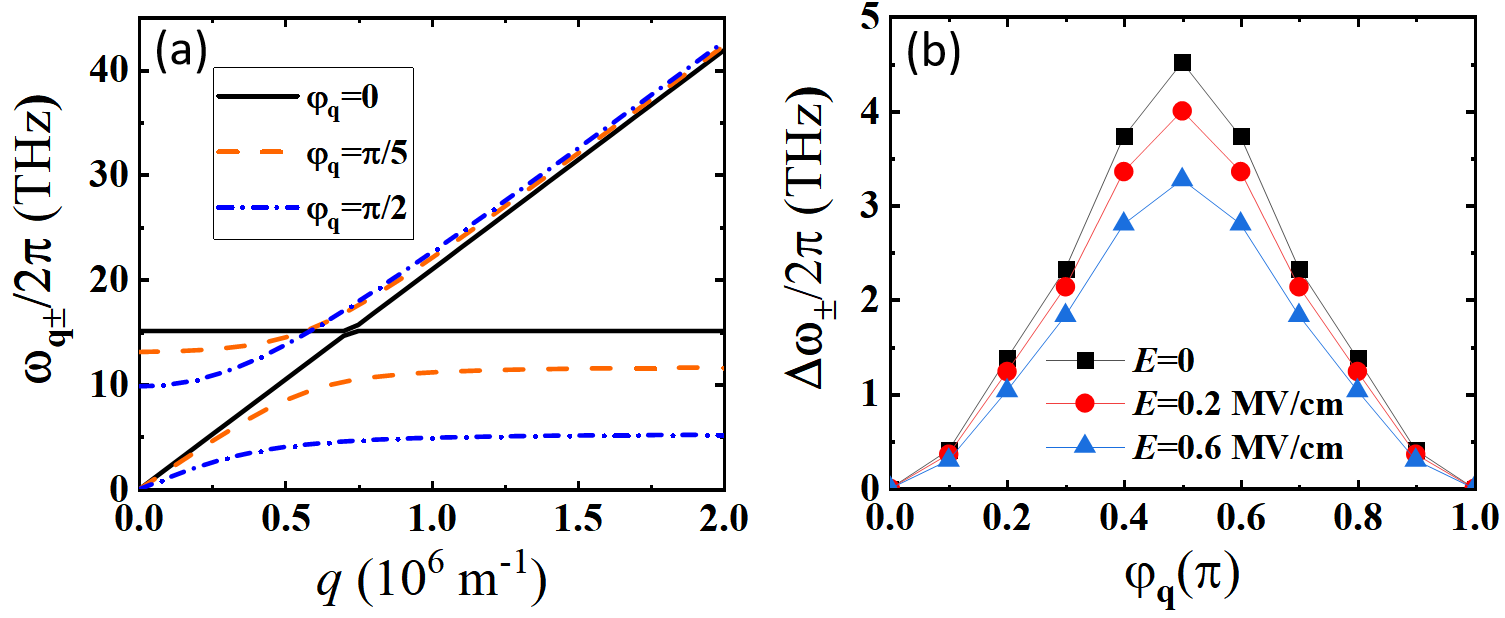}\newline\caption{(a) The ferron polariton dispersion relation for different angles \(\varphi_{\mathbf{q}}\) between wave vector and the polar axis. (b) The anti-crossing gap between the upper and lower branches plotted as a function of \(\varphi_{\mathbf{q}}\) is a measure of the ferron-photon coupling strength, while the dependence on an applied static field (\(E\)) along the polar axis measures the ferron dc dipole.}%
\label{Fig-polariton}%
\end{figure}

Fig.~\ref{Fig-polariton} shows the computed $\omega_{\mathbf{q}\pm}$ as a function of wave vector at different directions. In contrast to conventional (isotropic) phonon polariton \cite{Born1956,Fano1956,Henry1965}, the ferron polariton gap exhibits both a strong direction and static field dependence. When $\mathbf{q} \parallel \hat{\mathbf{z}}$ (i.e., $\varphi_{\mathbf{q}}=0$) a transversely polarized photon decouples from a ferron that is purely longitudinally polarized ($\mathbf{q}\parallel \mathbf{P}$), such that photon and ferron branches cross. The ferron-photon coupling emerges when $\mathbf{q}\nparallel\hat{\mathbf{z}}$; the ferron is then a mixed longitudinal and transverse wave that leads to an anti-crossing gap that increases with $\phi_{\mathbf{q}}$ and peaks at $\mathbf{q}\perp \hat{\mathbf{z}}$ (i.e., $\varphi_{\mathbf{q}}=\pi/2$). Because of the dc dipole moment of ferrons, the polariton gap can be tuned by a static field along the polar axis, which enables the modulation of THz radiation. In the previous work on displacive ferroelectrics \cite{Tang2022b}, a radiation field polarized along the polar axis was assumed, such that the strong anisotropy of the hybridization gap was missed.  

\emph{Thermal pumping of ferrons.---}Thermal ferrons reduce the total polarization by the time-independent term in Eq.~(\ref{Pol2}). The Boltzmann equation governs the thermal ferron distribution function $f_{\mathbf{q}}(\mathbf{r},t)$ when perturbed by, e.g., a spatiotemporal temperature field $T_{\text{ext}}(\mathbf{r}, t)$, i.e.,
\begin{equation}
\frac{\partial f_{\mathbf{q}}}{\partial t}+\mathbf{v}_{\mathbf{q}}\cdot\boldsymbol{\nabla} f_{\mathbf{q}}=-\frac{f_{\mathbf{q}}-n_{P}(\mathbf{r},t)}{\tau_{N}}-\frac{f_{\mathbf{q}}-n_P^{\ast}(\mathbf{r},t)}{\tau_{C}}, \label{BOLz}
\end{equation} 
where $\mathbf{v}_{\mathbf{q}}=\partial\omega_{\mathbf{q}}/\partial\mathbf{q}$ the ferron group velocity. The collision term on the right-hand side of Eq.~(\ref{BOLz}) represents the scatterings by ferron-ferron interactions, defects, electrically inactive phonons, etc. Analogous to the magnon-diffusion theory \cite{Zhang2012, Cornelissen2016}, here we adopt a dual relaxation time approximation for ferron number-conserving and -non-conserving scattering processes with relaxation times $\tau_{C}$ and $\tau_{N}$, respectively. The $\tau_N$-term relaxes the distribution towards the local Planck distribution $n_{P}(\mathbf{r},t)=[\exp(\hbar\omega_{\mathbf{q}}/k_{B}T_{\text{ext}}(\mathbf{r},t))-1]^{-1}$, while the $\tau_{C}$-term favors an intermediate Bose-Einstein distribution $n_{P}^{\ast}(\mathbf{r},t)=\{\exp[(\hbar\omega_{\mathbf{q}}-\mu(\mathbf{r},t))/k_{B}T_{\text{ext}}(\mathbf{r},t)]-1\}^{-1}$ with a quasi-equilibrium chemical potential $\mu(\mathbf{r},t)$, assuming that thermalization to the local external temperature is fast and that \(\tau_C \ll \tau_N\).  

When $T_{\text{ext}}(\mathbf{r},t)$ and $\mu(\mathbf{r}, t)$ vary smoothly and slowly on the scale of the ferron mean free path and the relaxation times, the solution to Eq.~(\ref{BOLz}) $f_{\mathbf{q}}=\tau[1+\tau(\partial_{t}+\mathbf{v}_{\mathbf{q}}\cdot\boldsymbol{\nabla})]^{-1}[\tau_{N}^{-1} n_{P}+\tau_{C}^{-1} n_{P}^{\ast}]$, where $\tau=(\tau_{C}^{-1}+\tau_{N}^{-1})^{-1}$, may be expanded in terms of powers of $\tau(\partial_{t}+\mathbf{v}_{\mathbf{q}}\cdot\boldsymbol{\nabla})$. To leading order, the condition for the ferron-conserving scattering $\sum_{\mathbf{q}}\tau_{C}^{-1}[f_{\mathbf{q}}-n_{P}^{\ast}(\mathbf{r},t)]=0$ leads to the diffusion equation  
\begin{equation}
(\partial_{t}-D_{f}\boldsymbol{\nabla}^{2}+\tau_{f}^{-1})\mu(\mathbf{r},t)=(g_{s}\boldsymbol{\nabla}^2-g_{t}\partial_{t} )T_{\text{ext}}(\mathbf{r},t), \label{Diffu}
\end{equation}where $D_{f}=\tau \overline{v_{\mathbf{q}}^{2}}/3$ is the diffusion coefficient, $\tau_{f}=\tau_{N}^2/(\tau_{N}+\tau_{C})$ the effective ferron number decay time, $g_{t}= (\tau_{C}/\tau)\overline{\hbar\omega}_{\mathbf{q}}/T_{0} $, $g_{s}=\tau_{C}(\overline{\hbar\omega_{\mathbf{q}}v_{\mathbf{q}}^2})/(3T_{0})$, and $T_{0}$ is a uniform background temperature. Here the overbar indicates the average $\overline{O_{\mathbf{q}}}\equiv\sum_{\mathbf{q}} O_{\mathbf{q}}(\partial n_{P}/\partial\omega_{\mathbf{q}})/\sum_{\mathbf{q}}(\partial n_{P}/\partial\omega_{\mathbf{q}}) $, and we simplified the notation by dropping the dependence on the transport direction. Eq.~(\ref{Diffu}) resembles the spin diffusion equations in metals \cite{Valet1993} and insulators \cite{Zhang2012,Cornelissen2016} and its form does not depend on the microscopic model and holds for all type of ferroelectrics \cite{Bauer2021,Adachi2023}. 
 
 We now illustrate the notion of thermal ferron pumping by a specific device consisting of a thin perpendicularly (out-of-plane) polarized ferroelectric film in $xy$ plane with an attached metallic wire of width $w$ along the $y$ direction. The Joule heating by a current in the wire causes a temperature distribution $T_{\text{ext}}(x, t)$ in the ferroelectric that obeys the one-dimensional heat diffusion equation $(C_{v}\partial_{t}-\kappa\nabla^2) T_{\text{ext}}(x,t)=Q(x,t) $, with $C_{v}$ and $\kappa$ the specific heat per unit volume and the thermal conductivity of the ferroelectric, respectively. $Q(x,t)=A(t)/(\sqrt{2\pi}w)\exp[-(x-x_{0})^2/2w^2]$ is the heat production rate under the wire, with amplitude $A(t)$ and location $x_{0}$ (i.e., $Q(x, t)\to A(t)\delta (x-x_{0})$ when $w\rightarrow 0$). Such a profile may be generated as well by the absorption of a line of focused laser light. When $A(t)$ varies slowly compared to the ferron relaxation times, we may discard the time derivative and obtain 
\begin{equation}
\mu(x, t)\simeq-\frac{g_{s}\lambda_{f}}{2\kappa D_{f}}A(t)\exp\left(-\frac{\vert x-x_{0}\vert}{\lambda_{f}} \right)+\mathcal{O}(\partial_{t} A) \label{FerronChe}
\end{equation}
for $\lambda_{f}\equiv\sqrt{D_{f}\tau_{f}}\gg w$. A non-zero $\mu(x, t)$ reflects a non-equilibrium accumulation of the ferron number and associated electric polarization. The latter is detectable by a thermovoltage generated by the associated surface-bound charge using, e.g., the electrostatic (Kelvin) force microscopy technique \cite{Girard2001}. Leading non-adiabatic terms cause transient screening charge currents in a metallic contact ($\partial_{t}\mu$) \cite{Bauer2021}. 

\begin{figure}
\centering
\par
\includegraphics[width=6.2cm]{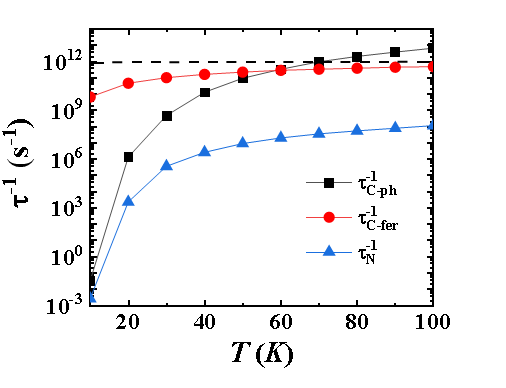}\newline\caption{The temperature dependence of ferron relaxation rates (defined in the text). The dotted horizontal line represents a temperature-independent scattering rate by the elastic impurities.}%
\label{Fig-relaxation}%
\end{figure}

\emph{Ferron relaxation times.---}Ferrons may be scattered by other ferrons, static disorder including rough sample boundaries, electrically inactive phonons, \textit{etc}. A quantitative theory depends on the specific material and sample shape and quality, which is beyond the scope of this paper. In the Supplemental Materials \cite{SM}, we estimate the relaxation rates by considering the scattering events by defects, acoustic and optical phonons, as well as three- and four-ferron interactions from the cubic and quartic terms in the Holstein-Primakoff expansion. $\tau_{N}^{-1}$ is contributed by the ferron-to-optical-phonon conversion and three-ferron processes, while $\tau_{C}^{-1}$ by the defects $\tau_{i}^{-1}$, ferron-ferron scatterings $\tau_{C-\text{fer}}^{-1}$, and the scatterings by acoustic phonons $\tau_{C-\text{ph}}^{-1}$. Fig.~\ref{Fig-relaxation} presents the temperature dependence of the average relaxation rates by each mechanism, showing that $\tau_{N}^{-1}\ll\tau_{C}^{-1}=\tau_{i}^{-1}+\tau_{C-\text{fer}}^{-1}+\tau_{C-\text{ph}}^{-1}$ because of the limited phase space for the ferron-non-conserving scatterings \cite{SM}. At elevated temperatures, $\tau_{C-\text{fer}}^{-1}$ ($\tau_{C-\text{ph}}^{-1}$) scales with temperature as $T^{2}$ (\(T\)). For a temperature-independent defect-induced $\tau_{i}\sim $ps and a ferron group velocity $\mathcal{O}(\mathrm{km/s})$, the ferron decay length ranges from tens of nanometers to centimeters for temperatures from 100 to 10\,K, implying efficient transport of electric polarization by ferrons.  

\emph{Conclusion and outlook.---}We analyze the properties of ferron excitations in order-disorder ferroelectrics by a microscopic pseudo-spin model with both ac and dc electric dipoles. The ferron gap can be tuned by a moderate electric field. Ferrons may be pumped parametrically with a threshold that allows the measurement of their dipole moment. Ferron-photon hybrid states display anti-crossing gaps that depend strongly on their propagation direction and the gate field. We also address the diffuse transport carried by ferrons under temperature gradients. Although focusing on order-disorder ferroelectrics, our results are generic and should describe the ferron excitations in other types of ferroelectrics, but model parameters may be very different. 

\emph{Acknowledge.---}We acknowledge support by JSPS KAKENHI Grants (Nos. 19H00645, 22H04965 and K23K130500).


\begin{thebibliography}{99}
\bibitem{Notedip}The dipole moment of the ions or charged units is given by their effective Born charge times off-center displacement.
\end{thebibliography}
\end{document}


\title{Supplementary Material to \textquotedblleft Electric Analog of Magnon Excitations in Order-Disorder Ferroelectrics\textquotedblright}
\author{Ping Tang$^{1}$}
\author{Gerrit E. W. Bauer$^{1,2,3,4,5}$}
\affiliation{$^1$WPI-AIMR, Tohoku
University, 2-1-1 Katahira, Sendai 980-8577, Japan}
\affiliation{$^2$Institute for Materials Research, Tohoku University,
2-1-1 Katahira, Sendai 980-8577, Japan} 
\affiliation{$^3$Center for Science and Innovation in Spintronics (CSIS), Tohoku University, Sendai 980-8577, Japan}
\affiliation{$^5$Kavli Institute for Theoretical Sciences, University of the Chinese Academy of Sciences,
Beijing 10090, China}
\maketitle

\begin{figure}
\centering
\par
\includegraphics[width=12cm]{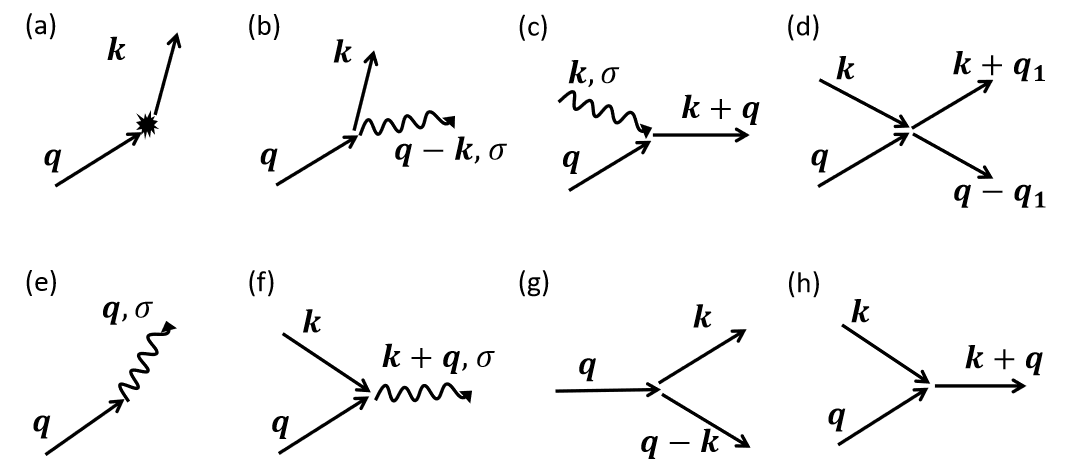}\newline\caption{Diagrams illustrating relaxation channels of ferrons with wave vector $\mathbf{q}$ indicated by solid arrows. The wavy lines are normal phonons with polarization \(\sigma\). (a) defect or boundary scattering, (b) and (c) ferron-conserving scattering by phonons, (d) four-ferron scattering, (e) one-ferron-to-phonon and (f) two-ferron-to-phonon conversion, and three-ferron splitting (g) and confluence (h).  }%
\label{Fig-scat}%
\end{figure}

In this Supplemental Material, we estimate the relaxation times of ferrons due to scattering at static impurities, phonons, and other ferrons, as schematically shown in Fig.~\ref{Fig-scat}. 

We consider three processes that contribute to the ferron scattering by the operator
\begin{equation}
    \hat{V}=\hat{V}_{i}+\hat{V}_{pl}+\hat{V}_{int}.
\end{equation}
$\hat{V}_{i}$ is the scattering potential caused by defects or boundaries that break translational symmetry.
\begin{equation}
\hat{V}_{i}=\sum_{\mathbf{q}^{\prime}}v_{\mathbf{q}^{\prime}\mathbf{q}}\hat{b}_{\mathbf{q}^{\prime}}^{\dagger}\hat{b}_{\mathbf{q}} ,
\end{equation}
where $v_{\mathbf{q}^{\prime}\mathbf{q}}$ a scattering amplitude, which in the limit of a single band conserves the ferron number, but not the momentum or the electric dipole, with temperature-independent ferron number-conserving relaxation rate 
\begin{align}
 \tau_{i}^{-1}&=\frac{2\pi}{\hbar}\sum_{\mathbf{q}^{\prime}}\vert v_{\mathbf{q}^{\prime}\mathbf{q}}\vert^2\delta(\hbar\omega_{\mathbf{q}}-\hbar\omega_{\mathbf{q}^{\prime}})\nonumber\\
&=\frac{2\pi}{\hbar} V^2 \mathcal{D}(\hbar\omega).
\end{align}
The second line holds for point-like scatterers and \(\mathcal{D}\) is the ferron density of states.

The electric polarization interacts with either the electric field created by optical polar phonons or acoustic phonons through the piezoelectric coupling by a pseudo-spin-lattice coupling of the form $\hat{V}_{pl}=\sum_{\mathbf{q}s} F_{\mathbf{q}s}\hat{S}_{-\mathbf{q}}^{z}\hat{Q}_{\mathbf{q}s}$ \cite{Blinc1972}, where $Q_{\mathbf{q}s}$ is the normal coordinate of a phonon mode $\mathbf{q}s$ and $F_{\mathbf{q}s}$ the coupling strength. In the long wavelength limit $NF_{\mathbf{q}O}^2\approx 10^{-5} -10^{-6}\,$(eV)$^3$ for the optical phonons while $F_{\mathbf{q}A}\rightarrow q$ for the acoustic phonons \cite{Kobayashi1968, Blinc1972}. In second quantization, the associated ferron-phonon coupling reads
\begin{align}
\hat{V}_{pl}=&-\frac{\sin\theta}{2}\sum_{\mathbf{q}s}F_{\mathbf{q}s}\sqrt{\frac{\hbar}{2\omega_{\mathbf{q}s}^{(ph)}}} (\hat{c}_{\mathbf{q}s}+\hat{c}_{-\mathbf{q}s}^{\dagger})(\hat{a}_{\mathbf{q}}^{\dagger}+\hat{a}_{-\mathbf{q}})-\frac{\cos\theta}{\sqrt{N}}\sum_{\mathbf{k}\mathbf{q}s}F_{\mathbf{q}s}\sqrt{\frac{\hbar}{2\omega_{\mathbf{q}s}^{(ph)}}}(\hat{c}_{\mathbf{q}s}+\hat{c}_{-\mathbf{q}s}^{\dagger}) \hat{a}_{\mathbf{k}}^{\dagger}\hat{a}_{\mathbf{k}-\mathbf{q}}\nonumber\\
&+\frac{\sqrt{N}}{2}\cos\theta \sum_{\mathbf{q}s}F_{\mathbf{q}s}\hat{Q}_{\mathbf{q}s}\delta_{\mathbf{q},0} \label{PL}
\end{align}
where $\hat{c}_{\mathbf{q}s}$ ($\hat{c}_{\mathbf{q}s}^{\dagger}$) is the annihilation (creation) operator of a phonon with frequency $\omega_{\mathbf{q}s}^{(ph)}$. The first bilinear term contributes to the hybridization sketched by (e) in Fig.~\ref{Fig-scat} that does not conserve the ferron number \cite{Blinc1972}, which is analogous to the magnetoelastic magnon  polarons \cite{Kikkawa2016, Flebus2017}. The second term represents the scattering by a phonon that conserves the number of ferrons $\hat{a}_{\mathbf{q}}^{\dagger}\hat{a}_{\mathbf{q}}$ but not $\hat{b}_{\mathbf{q}}^{\dagger}\hat{b}_{\mathbf{q}}$, while the last term describes ferroelasticity, \textit{i.e.} a static lattice deformation by the spontaneous ferroelectric order that does not affect the dynamics. The Eq.~(\ref{PL}) leads to the ferron-conserving and -nonconserving relaxation rates as
\begin{align}
\tau_{C,\text{ph}}^{-1}=&\frac{\pi\cos^2\theta}{ N}\sum_{\mathbf{q}^{\prime}s}\frac{\vert F_{\mathbf{q}^{\prime}s}\vert^2}{\omega_{\mathbf{q}^{\prime}s}^{(ph)}} 
\left[(n_{\mathbf{q}^{\prime}s}^{(ph)}-n_{\mathbf{q}-\mathbf{q}^{\prime}})\delta (\hbar\omega_{\mathbf{q}}-\hbar\omega_{\mathbf{q}-\mathbf{q}^{\prime}}+\hbar\omega_{\mathbf{q}^{\prime}s}^{(ph)})\right.\nonumber\\
&\left.+(n_{\mathbf{q}-\mathbf{q}^{\prime}}+n_{\mathbf{q}^{\prime}s}^{(ph)}+1)\delta (\hbar\omega_{\mathbf{q}}-\hbar\omega_{\mathbf{q}-\mathbf{q}^{\prime}}-\hbar\omega_{\mathbf{q}^{\prime}s}^{(ph)})\right]\\
\tau_{N,\text{ph}}^{-1}=&\frac{\pi}{4}\sin^{2}\theta\sum_{\mathbf{q}^{\prime}s}\frac{\vert F_{\mathbf{q}^{\prime}s} \vert^2}{\omega_{\mathbf{q}^{\prime}s}^{(ph)}}\delta(\hbar\omega_{\mathbf{q}}-\hbar\omega_{\mathbf{q}^{\prime}s}^{(ph)}).
\end{align}

The non-parabolicity of the local potentials of spins, atomic masses, and dipoles in solids can be interpreted in terms of interactions between the normal excitation modes. Here we focus on the 3- and 4-ferron scattering events in a single soft-phonon band by retaining the cubic and quartic terms in the Holstein-Primakoff expansion of the pseudo-spin operators, which gives
\begin{align}
\hat{V}_{\text{int}}=&-\frac{\sin2\theta}{4\sqrt{N}}\sum_{\mathbf{k}\mathbf{q}} (J_{\mathbf{q}}\hat{a}_{\mathbf{k}+\mathbf{q}}^{\dagger} \hat{a}_{\mathbf{k}} \hat{a}_{\mathbf{q}} +H.c.)+\frac{\sin^2\theta}{8N}\sum_{\mathbf{k}\mathbf{k}^{\prime}\mathbf{q}}(J_{\mathbf{k}}\hat{a}_{\mathbf{k}+\mathbf{q}}^{\dagger} \hat{a}_{\mathbf{q}-\mathbf{k}^{\prime}}\hat{a}_{\mathbf{k}^{\prime}}\hat{a}_{\mathbf{k}}+H.c.) \nonumber\\&+\sum_{\mathbf{k}\mathbf{k}^{\prime}\mathbf{q}}\left[\frac{\sin^2\theta}{8N}(J_{\mathbf{k}}+J_{\mathbf{k}+\mathbf{q}})-\frac{\cos^2\theta}{2N} J_{\mathbf{k}+\mathbf{q}-\mathbf{k}^{\prime}}\right]\hat{a}_{\mathbf{k}+\mathbf{q}}^{\dagger} \hat{a}_{\mathbf{k}^{\prime}-\mathbf{q}}^{\dagger}\hat{a}_{\mathbf{k}^{\prime}}\hat{a}_{\mathbf{k}} \label{Interaction}
\end{align}
where we discarded umklapp scattering. Since the canting angle $\theta$ ($\sim \Omega/J_{0}$) is usually small ($\leq 0.1$) , we may replace $\hat{a}_{\mathbf{q}}$ ($\hat{a}_{\mathbf{q}}^{\dagger}$) by the ferron operator of the normal mode $\hat{b}_{\mathbf{q}}$ ($\hat{b}_{\mathbf{q}}^{\dagger}$) in Eq.~(\ref{PL}) and Eq.~(\ref{Interaction}). Using Fermi’s Golden Rule to leading order in $\theta$
\begin{align}
\tau_{C,\text{fer}}^{-1}(\mathbf{q})=&\frac{1}{n_{\mathbf{q}}(n_{\mathbf{q}}+1)}\sum_{\mathbf{q}^{\prime}\mathbf{q}^{\prime\prime}\mathbf{q}^{\prime\prime\prime}} \frac{1}{2}Q_{\mathbf{q}\mathbf{q}^{\prime}}^{\mathbf{q}^{\prime\prime}\mathbf{q}^{\prime\prime\prime}} \label{tauC}\\
\tau_{N,\text{fer}}^{-1}(\mathbf{q})=&\frac{1}{n_{\mathbf{q}}(n_{\mathbf{q}}+1)}\sum_{\mathbf{q}^{\prime}\mathbf{q}^{\prime\prime}} \left(Q_{\mathbf{q}\mathbf{q}^{\prime}}^{\mathbf{q}^{\prime\prime}}+\frac{1}{2}Q_{\mathbf{q}}^{\mathbf{q}^{\prime}\mathbf{q}^{\prime\prime}}\right) \label{tauN}
\end{align}
with
\begin{align}
Q_{\mathbf{q}\mathbf{q}^{\prime}}^{\mathbf{q}^{\prime\prime}}=&Q_{\mathbf{q}}^{\mathbf{q}^{\prime}\mathbf{q}^{\prime\prime}}=\frac{\pi \theta ^2}{2\hbar N}(J_{\mathbf{q}}+J_{\mathbf{q}^{\prime}})^2 n_{\mathbf{q}}n_{\mathbf{q}^{\prime}}(n_{\mathbf{q}^{\prime\prime}}+1)\delta_{\mathbf{q}+\mathbf{q}^{\prime}, \mathbf{q}^{\prime\prime}} \delta(\hbar\omega_{\mathbf{q}}+\hbar\omega_{\mathbf{q}^{\prime}}-\hbar\omega_{\mathbf{q}^{\prime\prime}})\\
Q_{\mathbf{q}\mathbf{q}^{\prime}}^{\mathbf{q}^{\prime\prime}\mathbf{q}^{\prime\prime\prime}}=&\frac{2\pi}{N^2\hbar}(J_{\mathbf{q}-\mathbf{q}^{\prime\prime}}+J_{\mathbf{q}-\mathbf{q}^{\prime\prime\prime}})^2 n_{\mathbf{q}}n_{\mathbf{q}^{\prime}}(n_{\mathbf{q}^{\prime\prime}}+1)(n_{\mathbf{q}^{\prime\prime\prime}}+1)\delta_{\mathbf{q}+\mathbf{q}^{\prime},\mathbf{q}^{\prime\prime}+ \mathbf{q}^{\prime\prime\prime}}\delta(\hbar\omega_{\mathbf{q}}+\hbar\omega_{\mathbf{q}^{\prime}}-\hbar\omega_{\mathbf{q}^{\prime\prime}}-\hbar\omega_{\mathbf{q}^{\prime\prime\prime}}).
\end{align}
Here $Q_{\mathbf{q}\mathbf{q}^{\prime}}^{\mathbf{q}^{\prime\prime}}$ and $Q_{\mathbf{q}^{\prime\prime}}^{\mathbf{q}\mathbf{q}^{\prime}}$ are the scattering rates of 3-ferron confluence and splitting processes, respectively, while $Q_{\mathbf{q}\mathbf{q}^{\prime}}^{\mathbf{q}^{\prime\prime}\mathbf{q}^{\prime\prime\prime}}$ those of the 4-ferron scattering that conserves the ferron number. The factor $1/2$ accounts for the double counting in the wave vector sums.

The total ferron-conserving and nonconserving relaxation rates are 
\begin{align}
\tau_{C}^{-1}=&\tau_{i}^{-1}+\tau_{C,\text{ph}}^{-1}+\tau_{C,\text{fer}}^{-1}\\
\tau_{N}^{-1}=&\tau_{N,\text{ph}}^{-1}+\tau_{N,\text{fer}}^{-1}.
\end{align}
The ferron-nonconserving $\tau_{N}^{-1}$ is the sum of the ferron-to-phonon conversion ($\tau_{N,\text{ph}}^{-1}$) and three-ferron scattering $\tau_{N,\text{fer}}^{-1}$ rates. The former requires the intersection of the phonon and ferron bands, while the latter is allowed only for a particular ferron band structure that satisfies the energy conservation constraint. For the ferron dispersion derived in the main text, the lowest frequency ($\sim2\pi\cdot5.3\, \text{THz}$) exceeds the bandwidth ($\sim2\pi\cdot5.2\,\text{THz}$), so the three-ferron scattering processes are prohibited since $\omega_{\mathbf{k}_{1}}=\omega_{\mathbf{k}_{2}}+\omega_{\mathbf{k}_{3}}$ cannot be fulfilled. By adopting $N(\hbar F_{\text{TO}})^2\sim 2\times 10^{-6}\,$(eV)$^3$ for a transverse optical mode at $12\,\text{THz}$ \cite{Kobayashi1968} and $NF_{\text{TA}}^{2}/\omega_{\text{TA}}^2\sim 0.2\,$meV for a TA mode with velocity $\sim 1\,$km/s, we present the temperature dependence of the average relaxation rates by each mechanism in Figure~4 of the main text.